\documentclass{tMPH2e}

\usepackage[colorlinks=true,linkcolor=blue,citecolor=blue]{hyperref}

\begin{document}

\doi{10.1080/0026897YYxxxxxxxx}
 \issn{1362–3028}
\issnp{0026–8976}
\jvol{00}
\jnum{00} \jyear{2009} 


\title{Rydberg-Stark states in oscillating electric fields}

\author{V. Zhelyazkova and S. D. Hogan$^\ast$\thanks{$^\ast$Corresponding author. Email: s.hogan@ucl.ac.uk
\vspace{6pt}}\\\vspace{6pt} {\em{Department of Physics and Astronomy, University College London, Gower Street, London WC1E 6BT, U.K.}}\\\vspace{6pt}\received{v4.5 released September 2009} }

\maketitle

\begin{abstract}
Experimental and theoretical studies of the effects of weak radio-frequency electric fields on Rydberg-Stark states with electric dipole moments as large as 10000 D are reported. High-resolution laser spectroscopic studies of Rydberg states with principal quantum number $n=52$ and $53$ were performed in pulsed supersonic beams of metastable helium with the excited atoms detected by pulsed electric field ionisation. Experiments were carried out in the presence of sinusoidally oscillating electric fields with frequencies of 20~MHz, amplitudes of up to 120~mV/cm, and dc offsets of up to 4.4~V/cm.  In weak fields the experimentally recorded spectra are in excellent agreement with the results of calculations carried out using Floquet methods to account for electric dipole couplings in the oscillating fields. This highlights the validity of these techniques for the accurate calculation of the Stark energy level structure in such fields, and the limitations of the calculations in stronger fields where $n-$mixing and higher-order contributions become important.\bigskip

\begin{keywords}Rydberg states, Stark effect, radio-frequency electric fields
\end{keywords}

\end{abstract}

\section{Introduction}

Rydberg states of atoms and molecules with high principal quantum number, $n$,
play important roles in many areas at the interface between physics and physical chemistry.  For example, high Rydberg states are exploited in (i) spectroscopic studies of the structure and dynamics of molecular cations by high resolution zero kinetic energy photoelectron (ZEKE) spectroscopy~\cite{dethlefs84a,reiser88a,hollenstein01a}, (ii) the precise determination of ionisation and dissociation energies~\cite{liu09a,sprecher11a}, and (iii) investigations of energy transfer and ionization in atomic and molecular scattering~\cite{dunning84a}, including scattering from surfaces~\cite{hill00a,lloyd05a,sashikesh13a}. Recently, the very large electric dipole moments, exceeding 1000 D for values of $n>16$, associated with high Rydberg states have given rise to the development of efficient methods for decelerating~\cite{yamakita04a,vliegen04a,hogan12a,lancuba14a}, guiding~\cite{lancuba13a,allmendinger14a,ko14a}, and trapping~\cite{hogan08a,hogan09a,seiler11a} translationally cold samples of state-selected Rydberg atoms and molecules. These techniques have enabled experimental investigations of the effects of collisional and radiative processes on the decay of high Rydberg states over timescales which were not previously possible~\cite{seiler11b}, and offer a route to improvements in spectroscopic and collision energy resolution in experiments in the areas listed above.

In addition to permitting efficient deceleration and electric trapping, the large electric dipole moments of high Rydberg states also make them very sensitive to electric fields. This sensitivity has been exploited for the precise measurement and cancellation of time-independent, dc, laboratory electric fields~\cite{osterwalder99a}. However, often the effects of weak time-dependent oscillating fields, resulting from the presence of electrical laboratory noise, the motion of samples in electric traps, or perturbations from charged particles (electrons, or ions), are neglected. As described here, fields of this kind can give rise to the broadening of, and the introduction of sidebands on, the optical transitions to Rydberg states with large electric dipole moments~\cite{autler56a}. Accurate characterization of the effects of such fields on the spectral intensity distributions can therefore allow these states to be exploited for sensitive detection of radio-frequency electric fields. These fields can also be used to modify the interaction potentials in collisions between pairs of Rydberg atoms or molecules, or collisions between a Rydberg atom or molecule and ground state samples (see, e.g.,~\cite{gallagher08a}).

Time-dependent, amplitude-modulated electric fields~\cite{pillet83,vandenHeuvell84,pillet87} have been employed previously to measure the polarizabilities of high Rydberg states with quadratic Stark shifts~\cite{zhang94a}, for studies of quantum systems in the semiclassical limit~\cite{spellmeyer97a,yoshida12a}, and in investigations of the electric dipole interactions between pairs of Rydberg atoms~\cite{vanditzhuijzen09a,tretyakov14a,zhelyazkova15a}. Here, we present the results of high-resolution laser spectroscopic studies of the effects of amplitude-modulated electric fields on Rydberg-Stark states of atomic helium which possess large electric dipole moments. In weak dc electric fields many of the states studied exhibit linear Stark energy shifts. The effects of amplitude modulation at avoided crossings close to the Inglis-Teller electric field, where manifolds of Rydberg-Stark states with values of $n$ which differ by $\pm1$ cross each other, have also been investigated. For a particular value of $n$, the Inglis-Teller electric field, $F_{\mathrm{IT}}$, is $F_{\mathrm{IT}}=2hc\,R/(3ea_0\,n^5)$, where $h$ is the Planck constant, $c$ is the speed of light in vacuum, $R$ is the Rydberg constant, and $e$ and $a_0$ are the electron charge and the Bohr radius, respectively~\cite{gallagher94a}. Comparisons are made between the experimentally recorded spectra and the results of calculations of the Stark energy level structure in the amplitude-modulated fields used in the experiments. These calculations were performed using Floquet methods~\cite{pillet87,shirley65a} to introduce sidebands on the unperturbed Rydberg states. The results are in excellent agreement with the experimental data for low-amplitude modulation, and highlight the limitations of the calculations for stronger modulation where higher-order contributions become important.

In the following, a description of the apparatus used in the experiments is first provided. This is then followed by a description of the Floquet methods used to calculate the energy level structure of the Rydberg-Stark states in the amplitude-modulated electric fields of the experiments. The results of the experiments and calculations for a range of dc offset electric fields and modulation amplitudes are then presented. 

\section{Experiment}

\begin{figure}
\begin{center}
\includegraphics[width=0.9\textwidth]{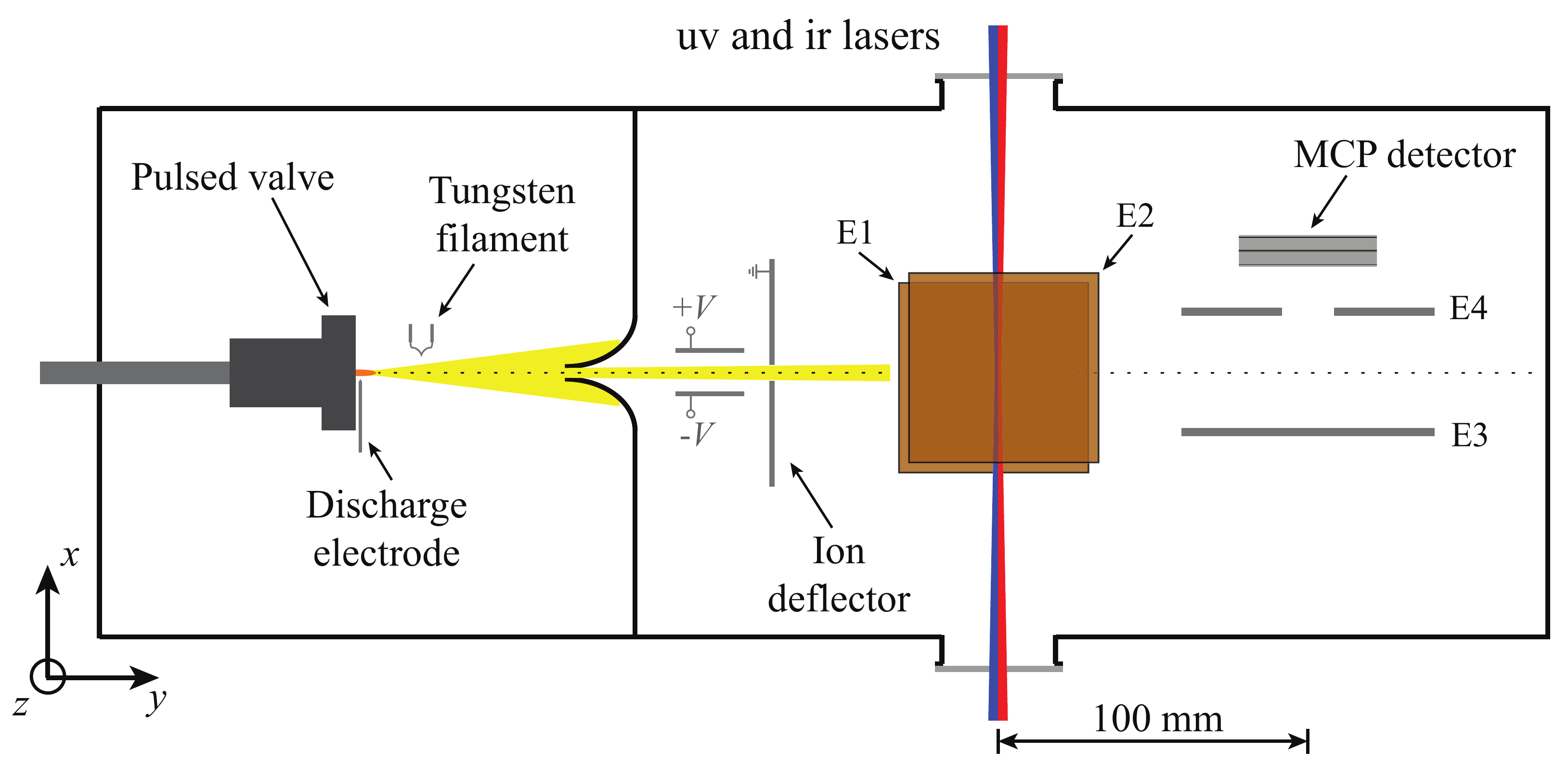}
\caption{Schematic diagram of the experimental apparatus (not to scale). The photoexcitation region is located between the two parallel copper plates labelled E1 and E2. Electric field ionisation of the excited Rydberg atoms occurs between the electrodes labelled E3 and E4 with the resulting ions accelerated toward the MCP detector.}%
\label{fig1}
\end{center}
\end{figure}

A schematic diagram of the apparatus used in the experiments is presented in Fig.~\ref{fig1}. A pulsed supersonic beam of helium atoms in the metastable 1s2s\,$^3$S$_1$ level is generated in a dc electric discharge at the exit of a pulsed valve~\cite{halfmann00a}. The valve is operated at a repetition rate of 50~Hz, a stagnation pressure of 6~bar, and with an opening time of 200~$\mu$s. The discharge occurs between a sharp metal anode, located $\sim1$~mm downstream from the exit of the valve, to which an electrical potential of $+270$~V is applied. For optimal shot-to-shot stability the discharge is seeded with electrons emitted from a hot tungsten filament positioned 15~mm beyond the anode. After passing through a 2~mm diameter copper skimmer, the pulsed atomic beam travels between a set of metal plates where an electric field of ~100~V/cm is generated to deflect any stray ions produced in the discharge. The beam then enters the photoexcitation region of the apparatus located between two parallel 70~mm$\times$70~mm copper plates (labelled E1 and E2 in Fig.~\ref{fig1}) separated in the $z$-dimension by 8.4~mm (see Fig.~\ref{fig1} for the definition of the axes). 

At the center point between these two electrodes the atomic beam is intersected at a right angle by two co-propagating, focused continuous wave (cw) laser beams which drive a resonant 1s2s\,$^3$S$_1\rightarrow$ 1s3p\,$^3$P$_2\rightarrow$ 1s$n$s/1s$n$d two-photon transition. The first step of this excitation sequence is driven by the frequency-doubled output of a cw diode laser. This transition lies in the ultraviolet (uv) region of the electromagnetic spectrum at 25708.587~cm$^{-1}$ ($\equiv388.975$~nm) and is driven with the laser power set to 70~$\mu$W, corresponding to an average intensity in the photoexcitation region of $\sim700$~mW/cm$^{2}$. The second photoexcitation step is driven by the amplified output of a second infra-red (ir) cw diode laser. The wavenumber at which this laser is operated is tuned within $\pm1$~cm$^{-1}$ of 12705.527~cm$^{-1}$ ($\equiv787.059$~nm) to excited Rydberg-Stark states for which $n=52$ or $n=53$. The power of this laser is set to 185~mW corresponding to an average intensity of $\sim14$~W/cm$^{2}$ in the photoexcitation region. The wavenumbers of the uv and ir lasers are stabilised and calibrated, respectively, using a high-resolution wavelength meter. Both laser beams are linearly polarised in the $z$ dimension. The spectral resolution in the experiment is 0.00043~cm$^{-1}$ ($\equiv13$~MHz), limited by the interaction time between the atomic beam and the 
laser radiation. 

After interacting with the lasers and exiting the photoexcitation region, the atomic helium beam  travels into the detection region of the apparatus. At a flight-time of 55~$\mu$s, when the atoms reach the center of the pair of copper plates labelled E3 and E4 in Fig.~\ref{fig1}, the electric potentials, initially both set to 0~V are rapidly switched with a pulse of $+1.4$~kV applied to E3. This generates an electric field of $\sim380$~V/cm to ionise the excited atoms, and accelerate the resulting He$^+$ ions through the aperture in E4 (which is grounded) and toward a microchannel plate (MCP) detector. The MCP signal resulting from this ion current is then integrated as a measure of the number of excited atoms. 

The dc and amplitude-modulated electric fields required for the experiments reported here were generated in the photoexcitation region by applying the appropriate electric potentials to electrodes E1 and E2. Stray dc electric fields in this region were compensated spectroscopically to $\pm1$~mV/cm and the absolute magnitudes of the applied fields were determined to the same precision.

\section{Floquet calculations}

The experiments reported here were performed in dc and time-dependent amplitude-modulated  electric fields acting in the $z$-dimension, i.e., $F(t) = (0,0,F_z (t))$. The $z$ component of this field can be expressed in the form
\begin{eqnarray}
F_z (t) = F_0 + F_{\mathrm{osc}}\,\cos(\omega_{\mathrm{rf}}t),\label{eq:Fz}
\end{eqnarray}
where $F_0$ is the magnitude of the dc offset field, and $F_{\mathrm{osc}}$ is the amplitude of the modulation applied at an angular frequency $\omega_{\mathrm{rf}}$. For all experiments and calculations presented $F_{\mathrm{osc}}\ll F_0$.

\subsection{Energy level structure in time-independent electric fields}\label{sec:timeindep}
In the absence of the oscillatory component of the electric field, the Rydberg-Stark energy level structure, and spectral intensities, can be accurately calculated by determining the eigenvalues and eigenvectors of the summed zero-field ($H_0$), and Stark ($H_{\mathrm{S}}$) Hamiltonian matrices~\cite{zimmerman79a,grimmel15}. This leads to the dc Hamiltonian
\begin{eqnarray}
H_{\mathrm{dc}} = H_0 + H_{\mathrm{S}}.\label{eq:H1}
\end{eqnarray}
The weak spin-orbit coupling in high Rydberg states means that $H_0$, constructed in an $|n\,\ell\,m_{\ell}\rangle$ basis, where $\ell$ and $m_{\mathrm{\ell}}$ are the orbital angular momentum quantum number and the azimuthal quantum number of the Rydberg electron, respectively, is diagonal in the absence of an external field. For each value of $n$ and $\ell$, the diagonal elements are given by the Rydberg formula including the appropriate quantum defects, $\delta_{\ell}$, for the non-hydrogenic low-$\ell$ sates. For the triplet Rydberg states of helium with values of $n$ close to 52 which are of interest here, the quantum defects of the s, p, d, f, g, and h states are $\delta_{\mathrm{s}}=0.296671$, $\delta_{\mathrm{p}}=0.068353$, $\delta_{\mathrm{d}}=0.002889$, $\delta_{\mathrm{f}}=0.000447$, $\delta_{\mathrm{g}}=0.000127$, and $\delta_{\mathrm{h}}=0.000049$~\cite{drake91}.

\begin{figure}
\begin{center}
\includegraphics[width=0.75\textwidth]{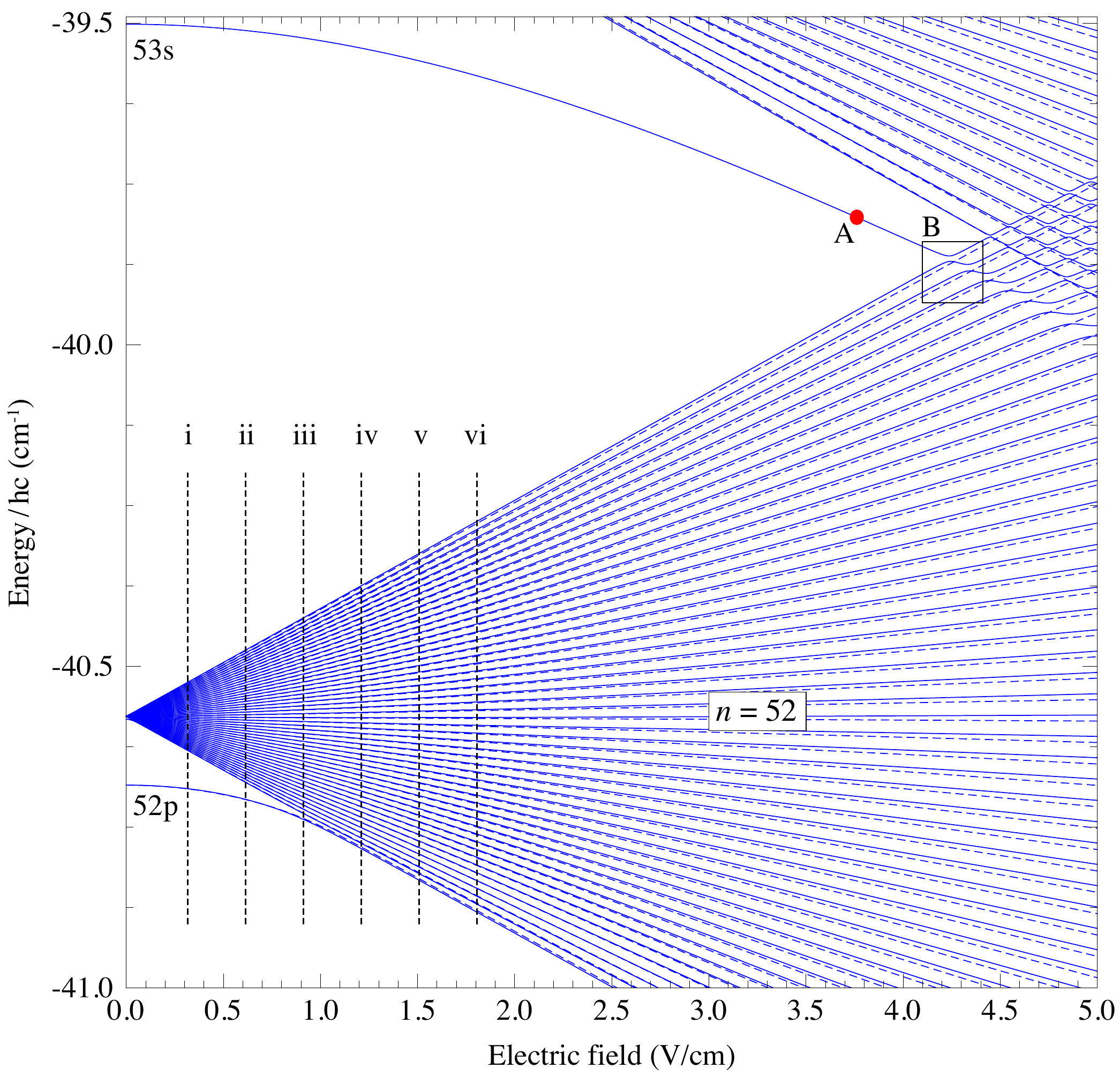}
\caption{Calculated Stark map of the $m_{\ell}=0$ (continuous curves) and $|m_{\ell}|=1$ (dashed curves) triplet Rydberg states of helium for $n=52$ and 53. The fields indicated by the dashed vertical lines (labelled i-vi), the dot labelled $A$, and the box labelled $B$ are discussed in the text. The energy on the vertical axis is defined with respect to the ionization limit.}
\label{fig2}
\end{center}
\end{figure}

The Stark Hamiltonian, $H_{\mathrm{S}}=eF_0\,z$ can be expressed in spherical polar coordinates as $H_{\mathrm{S}}=eF_0\,r\cos\theta$, and gives rise to matrix elements, coupling states $|n\,\ell\,m_{\ell}\rangle$ and $|n'\ell'\,m_{\ell}'\rangle$, of the form
\begin{eqnarray}
\langle n'\ell'm_{\ell}'|H_{\mathrm{S}}|n\,\ell\,m_{\ell}\rangle &=& \langle n'\ell'm_{\ell}'|eF_0\,r\cos\theta|n\,\ell\,m_{\ell}\rangle\\
&=& eF_0\langle \ell'm_{\ell}'|\cos\theta| \ell\,m_{\ell} \rangle\langle n'\ell'|r|n\,\ell\rangle.\label{eq:HS}
\end{eqnarray}
The angular components of this expression can be determined by expansion in terms of spherical harmonics. This leads to the result that the matrix elements vanish unless $m_{\ell}' = m_{\ell}$ and $\ell' = \ell\pm1$, and to the angular integrals~\cite{zimmerman79a} 
\begin{eqnarray}
\langle (\ell+1)\,m_{\ell}|\cos\theta| \ell\,m_{\ell} \rangle &=& \sqrt{\frac{(\ell+1)^2 - m_{\ell}^2}{(2\ell+3)(2\ell+1)}}\label{eq:angint1}
\end{eqnarray}
and
\begin{eqnarray}
\langle (\ell-1)\,m_{\ell}|\cos\theta| \ell\,m_{\ell} \rangle &=& \sqrt{\frac{\ell^2-m_{\ell}^2}{(2\ell+1)(2\ell-1)}}\label{eq:angint2}.
\end{eqnarray}
The radial matrix elements, $\langle n'\ell'|r|n\,\ell\rangle$, in Eq.~\ref{eq:HS} were calculated numerically using the Numerov method~\cite{zimmerman79a,gallagher94a}. 

Calculating the eigenvalues of the complete Hamiltonian, $H_{\mathrm{dc}}$ in Eq.~\ref{eq:H1}, for values of $n$ close to 52, over a range of electric fields from 0~V/cm to 5~V/cm, results in the $m_{\ell}=0$ and $|m_{\ell}|=1$ Stark maps displayed in Fig.~\ref{fig2}. For these calculations, all states for which $50\leq n\leq 54$ were included in the basis used to ensure convergence over the range of energies and electric field strengths encompassed in the figure.

\subsection{Energy level structure in time-dependent electric fields}\label{sec:timedep}
The approach, described in Section~\ref{sec:timeindep}, which was employed to calculate the Stark energy level structure in time-independent, dc electric fields can be extended to treat the effects of time-dependent, oscillating electric fields using the Floquet method~\cite{pillet87,shirley65a}. In the presence of the time-dependent electric field in Eq.~\ref{eq:Fz}, the complete Hamiltonian, $H_{\mathrm{mod}}(t)$, takes the form
\begin{eqnarray}
H_{\mathrm{mod}}(t) = H_0 + H_{\mathrm{S}} + H_{\mathrm{F}}(t),\label{eq:Hmod}
\end{eqnarray}
where within the dipole approximation
\begin{eqnarray}
H_{\mathrm{F}}(t)&=&eF_{\mathrm{osc}}\,\cos(\omega_{\mathrm{rf}}t)\,z\\
 &=&eF_{\mathrm{osc}}\,\cos(\omega_{\mathrm{rf}}t)\,r\cos\theta
 \end{eqnarray}
is the term arising from the oscillatory component of the field. Because $H_{\mathrm{mod}}(t)$ contains a term that is periodic in time, the solutions of the Schr\"odinger equation for the system described by this Hamiltonian must also be periodic in time and therefore the resulting wavefunctions must contain Fourier components occurring at the fundamental, and harmonics, of the modulation frequency. These Fourier components manifest themselves as sidebands on the Rydberg states.

The effects of the amplitude modulation of the electric field on the Rydberg states can be accounted for by extending the basis used for the calculations described in Section~\ref{sec:timeindep} to include these Floquet sidebands. Each sideband in the resulting $|n\ell m_{\ell}\,q\rangle$ basis is labelled by the integer index $q$, where $-q_{\mathrm{max}}\leq q\leq q_{\mathrm{max}}$, with $q_{\mathrm{max}}$ representing the maximum index included in the basis. The displacement in energy of sideband $q$ from each unperturbed Rydberg states is then $q\hbar\omega_{\mathrm{osc}}$. 

Because the dc and time-dependent components of the electric field in Eq.~\ref{eq:Fz} both act in the $z$ direction, neither mixes states with different values of $m_{\ell}$. As in the case discussed in Section~\ref{sec:timeindep}, in the Floquet basis the dc field only couples states for which $m_{\ell}'=m_{\ell}$, $q'=q$, and $\ell'=\ell\pm1$. The corresponding matrix elements are therefore those given by Eq.~\ref{eq:HS}. The time-dependent component of the field couples states for which $m_{\ell}'=m_{\ell}$, $q'=q\pm1$, and $\ell'=\ell\pm1$~\cite{spellmeyer98a}. The resulting matrix elements can be expressed in the form
\begin{eqnarray}
\langle n'\ell'm_{\ell}'\,q'|H_{\mathrm{F}}|n\,\ell\,m_{\ell}\,q\rangle &=& \delta_{q',q\pm1}\,\frac{1}{2}\langle n'\ell'm_{\ell}'\,q'|eF_{\mathrm{osc}}\,r\cos\theta|n\,\ell\,m_{\ell}\,q\rangle\\\nonumber\\
&=& \delta_{q',\,q\pm1}\,\frac{e\,F_{\mathrm{osc}}}{2}\langle \ell'm_{\ell}'|\cos\theta| \ell\,m_{\ell} \rangle\langle n'\ell'|r|n\,\ell\rangle,\label{eq:HF}
\end{eqnarray}
where $\delta_{x,y}$ is a Kronecker delta function. As in Eq.~\ref{eq:HS}, expansion of the angular components of this expression in terms of spherical harmonics leads to the result that these matrix elements are also only non-zero if $m_{\ell}'=m_{\ell}$ and $\ell'=\ell\pm1$. These angular integrals have the same form as those in Eqs.~\ref{eq:angint1} and~\ref{eq:angint2}.

The eigenvalues and eigenvectors of the complete Hamiltonian matrix, $H_{\mathrm{mod}}$ in Eq.~\ref{eq:Hmod}, describing the Rydberg states in the amplitude-modulated electric field with a dc offset were calculated by matrix diagonalisation. For all of the calculations described here a basis of Rydberg states ranging from $50\leq n\leq54$ was considered. Separate calculations were performed for basis states with $m_{\ell}=0$ and $|m_{\ell}|=1$. To achieve convergence in the calculated eigenvalues up to the Inglis-Teller limit at $n=52$ ($\sim$4.2~V/cm), for weak amplitude modulation, i.e., for $F_{\mathrm{osc}}\leq30$~mV/cm, values of $q$ up to $q_{\mathrm{max}}=8$ were required, while for stronger amplitude modulation, i.e., for $30$~mV/cm $\leq F_{\mathrm{osc}} \leq 120$~mV/cm, values of $q$ up to $q_{\mathrm{max}}=17$ were employed.

\subsection{Spectral intensities}
Using the resonant two-photon excitation scheme employed in the experiments described here, $n$s and $n$d Rydberg states were photoexcited from the metastable 1s2s\,$^3$S$_1$ level. Population of the $M_J=0, \pm1$ sublevels of the intermediate 1s3p\,$^3$P$_2$ level resulted from photoexcitation using uv laser radiation that was linearly polarised along the electric field axis ($z$-axis). A copropagating ir laser beam with the same linear polarisation was then used to drive transitions from these intermediate sublevels to the high Rydberg states. For an initially unpolarised population of atoms in the 1s2s\,$^3$S$_1$ level,  the first photoexcitation step results in 0.6 of the atoms being excited to the $|M_J|=1$ sublevels, and 0.4 of the atoms excited to the $|M_J|=0$ sublevel of the intermediate state.

The spectral intensities of transitions from the intermediate 1s3p\,$^3$P$_2$ level to each Floquet eigenstate with $n$s or $n$d character were calculated by first transforming from the spin-orbit coupled $|n L S J M_{J}\rangle$ basis to the $|n L M_{L} S M_{S}\rangle$ basis using the Clebsch-Gordon coefficients (see, e.g.,~\cite{zare88a}), where
\begin{eqnarray}
\langle L M_L\,S M_S|L S J M_J\rangle = (-1)^{-L+S-M_J}\sqrt{2J+1} \left( \begin{array}{ccc}
L & S & J \\
M_L & M_S & -M_J\end{array} \right).\label{eq:ClebschGordon}
\end{eqnarray}
For the triplet excited states of helium considered here, since $\ell=0$ and $m_{\ell}=0$ for the electron in the 1s orbital, $L$ and $M_L$ in Eq.~\ref{eq:ClebschGordon} are equivalent to $\ell$ and $m_{\ell}$, respectively, for the single excited electron. The Clebsch-Gordon coefficients for each permissible value of $m_{\ell}$, i.e. $m_{\ell}=0,\,\pm1$, of this intermediate state, together with the relative populations of each $M_J$ sublevel were then used to weight the dipole-allowed $\Delta m_{\ell}=0$ transitions to the Rydberg states. The spectral intensities of these transitions were determined using the coefficients, $C_{i,j}$, of the eigenvectors obtained following the calculations outlined in Sections~\ref{sec:timeindep} and~\ref{sec:timedep}. Here, $i$ is an index used to label the eigenstate of $H_{\mathrm{mod}}$, and $j$ is an index representing the $|n'_j,\ell'_j,m_{\ell}' ,q'_j\rangle$ basis state. By identifying the $q=0$ components of each eigenstate with $n$s or $n$d character, the spectral intensities, $S_{i}(m_{\ell}')$, associated with the set of allowed, $\Delta\ell=\pm1$ and $\Delta m_{\ell}=0$, electric dipole transitions from each $m_{\ell}$ sublevel of the intermediate 1s3p\,$^3$P$_{2}$ level are then
\begin{eqnarray}
S_{i}(m_{\ell}')&=& \left[\sum_{j} \delta_{q_j',q}\,C_{i,j}\,\langle n_j'\ell_j'\,m_{\ell}'\,q_j'|\mu_{z}Ê|n\,\ell\,m_{\ell}\,q\rangle\right]^2\\
&=& \left[\sum_{j} \delta_{q_j',0}\,C_{i,j}\,\langle \ell_j'\,m_{\ell}|\cos\thetaÊ|\ell=1\,m_{\ell}\rangle\langle n_j'\ell_j'|ea_0\,rÊ|n=3\,\ell=1\rangle\right]^2,\nonumber\\
\end{eqnarray}
where $m_{\ell}'=m_{\ell}=0,\pm1$. The angular integrals required in this calculation for transitions for which $m_{\ell}'=m_{\ell}$ and $\ell'=\ell\pm1$ are the same as those in Eqs.~\ref{eq:angint1} and~\ref{eq:angint2}. From this procedure it can be seen that the oscillatory component of the electric field introduces $q=0$ character into the Floquet sidebands with values of $|q|>0$. It is this $q=0$ character that gives these higher-order sidebands observable spectral intensity.

\section{Results}
To characterize the effects of the amplitude-modulated electric fields on the Rydberg-Stark states, we have carried out sets of experiments and calculations in a range of dc and amplitude-modulated fields, at various positions in the Stark map displayed in Fig.~\ref{fig2}. These include: (i) extended spectra for a range of dc offset fields but fixed values of $F_{\mathrm{osc}}$ and $\omega_{\mathrm{osc}}$, (ii) detailed studies of the effects of the modulation with a range of amplitudes, $F_{\mathrm{osc}}$, on an isolated Stark state with an electric dipole moment of 7900~D, and (iii) investigations of the effects of amplitude modulation at an avoided crossing in the Stark map at the Inglis-Teller limit between the $n=52$ and $n=53$ Stark manifolds. The results of these studies are discussed in the following.  

\subsection{Stark maps in dc and amplitude-modulated electric fields}
\begin{figure}[!h]
\begin{center}
\includegraphics[width=0.9\textwidth]{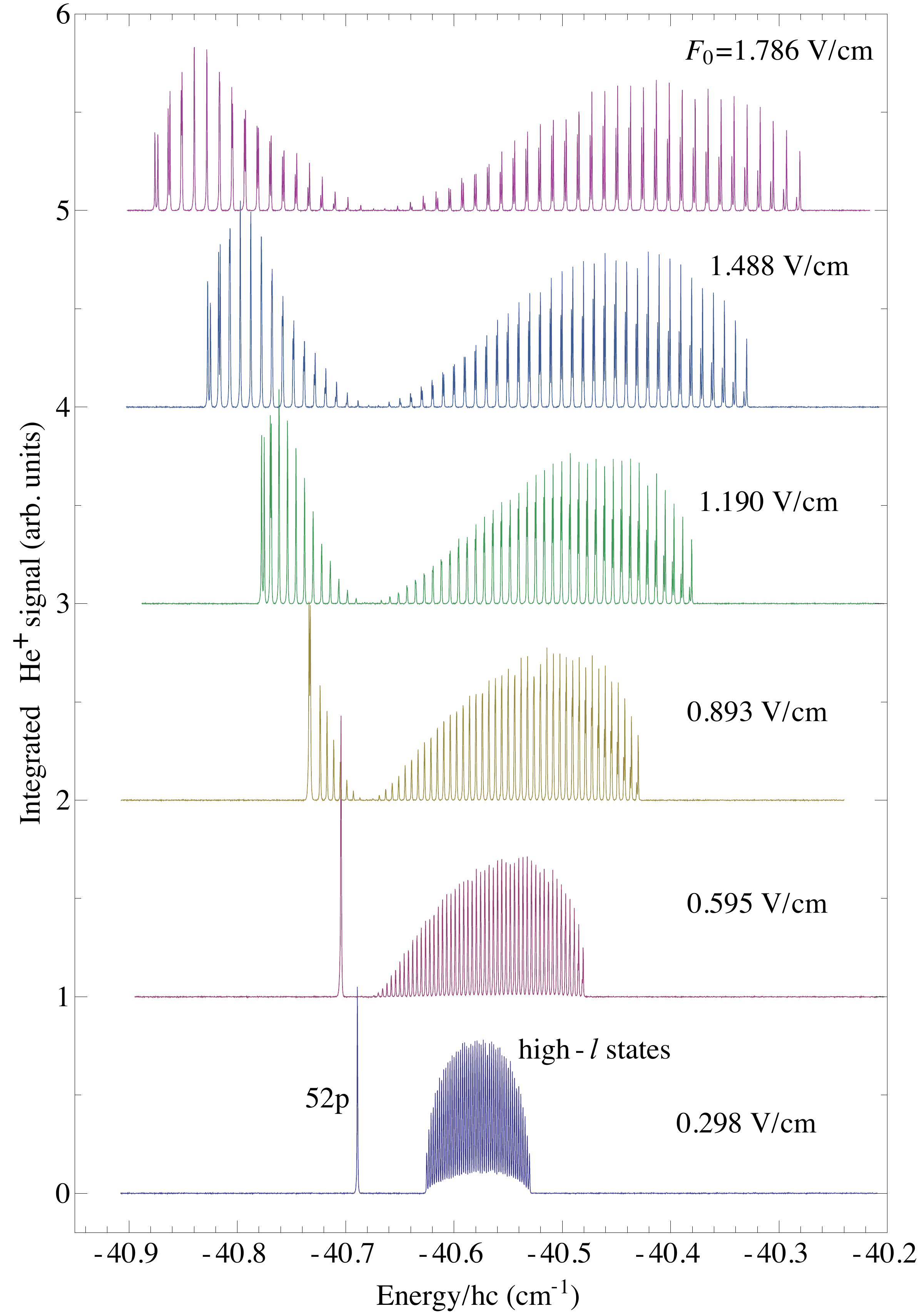}
\caption{Experimentally recorded spectra encompassing transitions to $n=52$ Rydberg states in dc electric fields of $F_0=0.298$, 0.595, 0.893, 1.190 and 1.786~V/cm (bottom to top). The leftmost (lowest energy) state evolves adiabatically to the 52p state in zero electric field. For clarity of presentation the upper spectra are vertically offset.}
\label{fig3}
\end{center}
\end{figure}

\begin{figure}[!h]
\begin{center}
\includegraphics[width=0.9\textwidth]{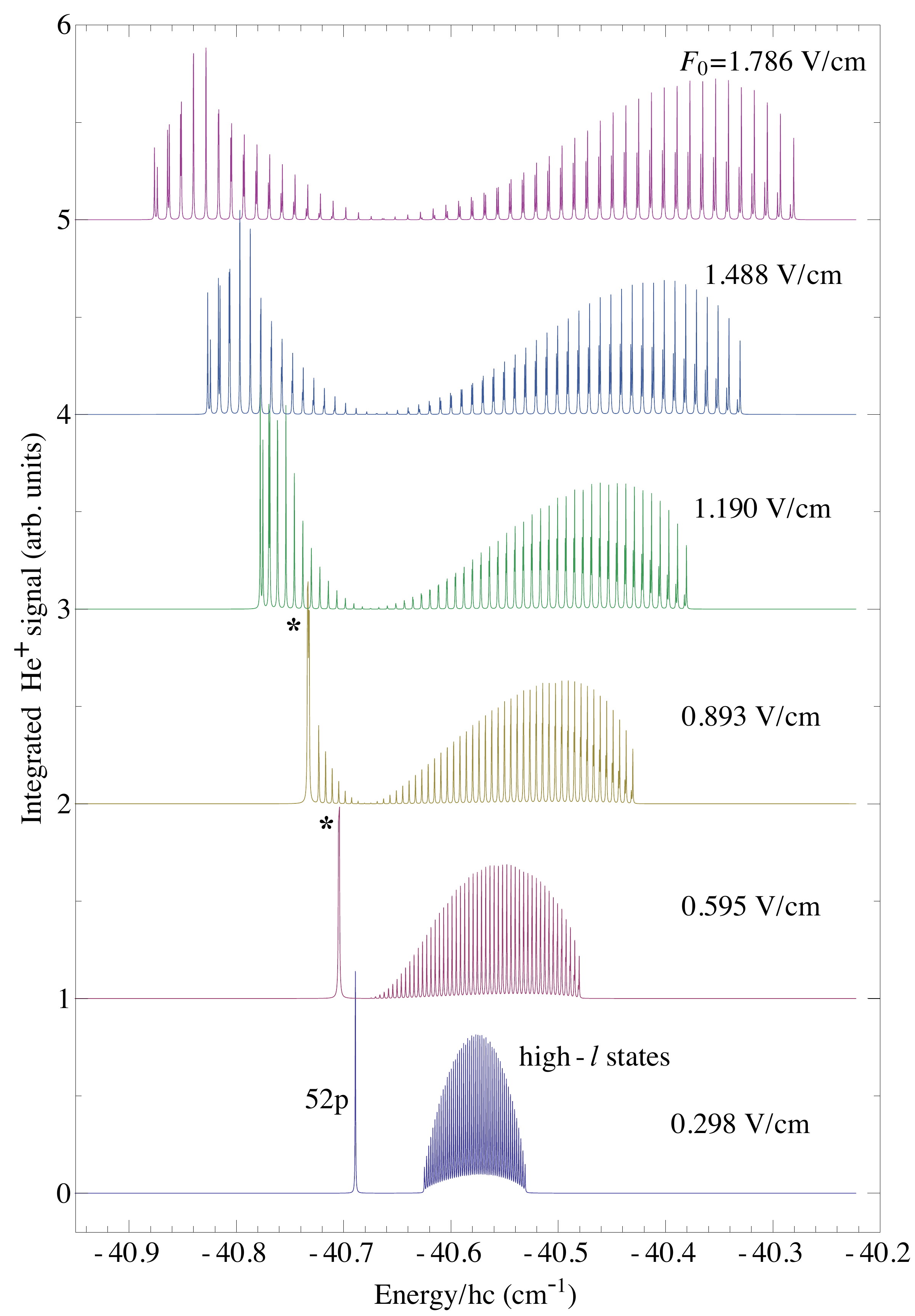}
\caption{Calculated spectra encompassing transitions to $n=52$ Rydberg states in dc electric fields of $F_0=0.298$, 0.595, 0.893, 1.190 and 1.786~V/cm (bottom to top). The leftmost (lowest energy) state evolves adiabatically to the 52p state in zero electric field. The asterisks indicate the two calculations in which the height of the field-free 52p state has been truncated to reflect the effects of saturation observed in the experiment (see text for details). For clarity of presentation the upper spectra are vertically offset.}
\label{fig4}
\end{center}
\end{figure}

\begin{figure}[!h]
\begin{center}
\includegraphics[width=0.9\textwidth]{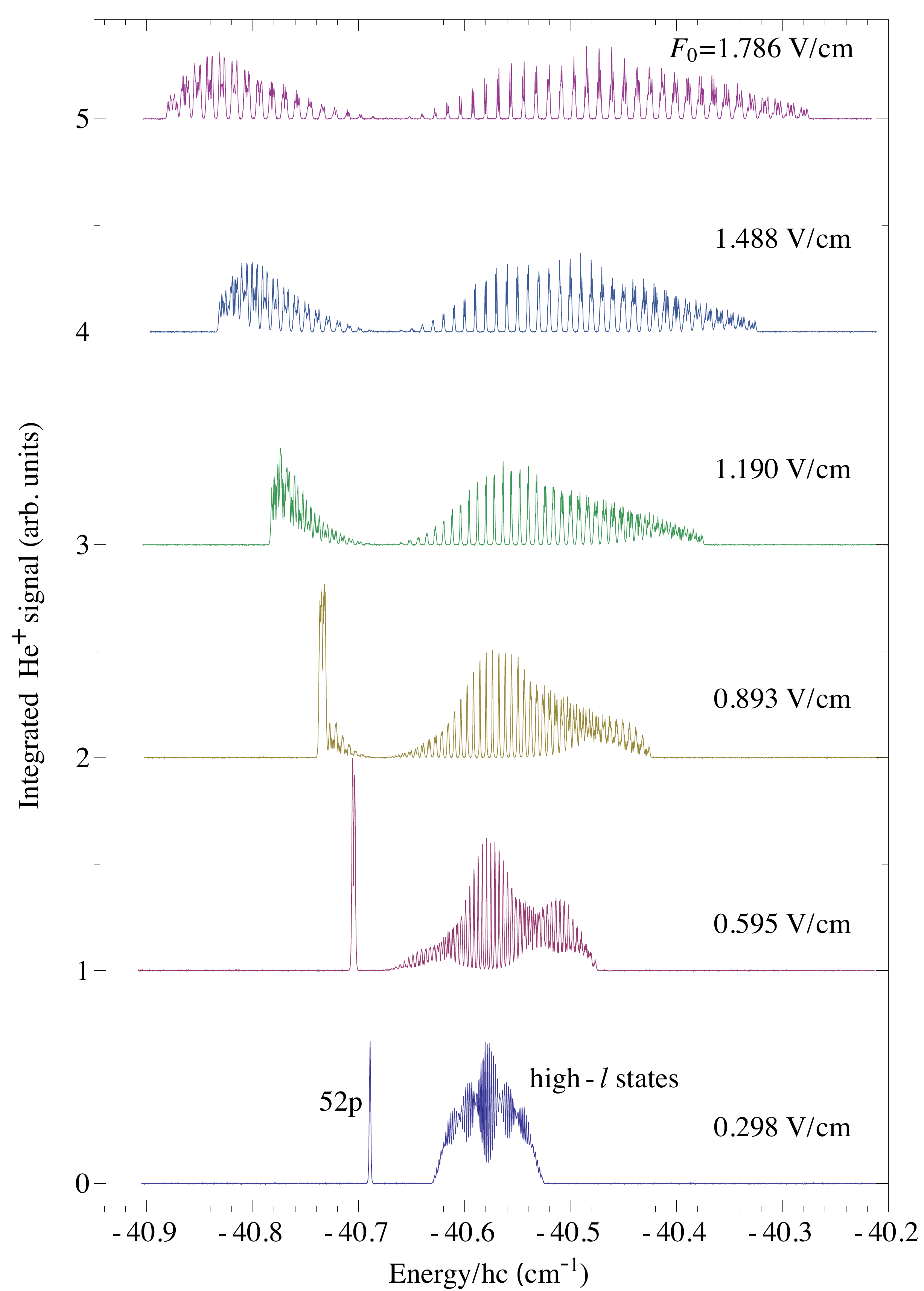}
\caption{Experimentally recorded spectra encompassing transitions to $n=52$ Rydberg states in electric field with dc offsets of $F_0=0.298$, 0.595, 0.893, 1.190 and 1.786~V/cm (bottom to top), and a modulation of amplitude $F_{\mathrm{osc}}=24.0$~mV/cm, and frequency $\omega_{\mathrm{rf}}=2\pi\times20$~MHz. For clarity of presentation the upper spectra are vertically offset.}
\label{fig5}
\end{center}
\end{figure}

\begin{figure}[!h]
\begin{center}
\includegraphics[width=0.9\textwidth]{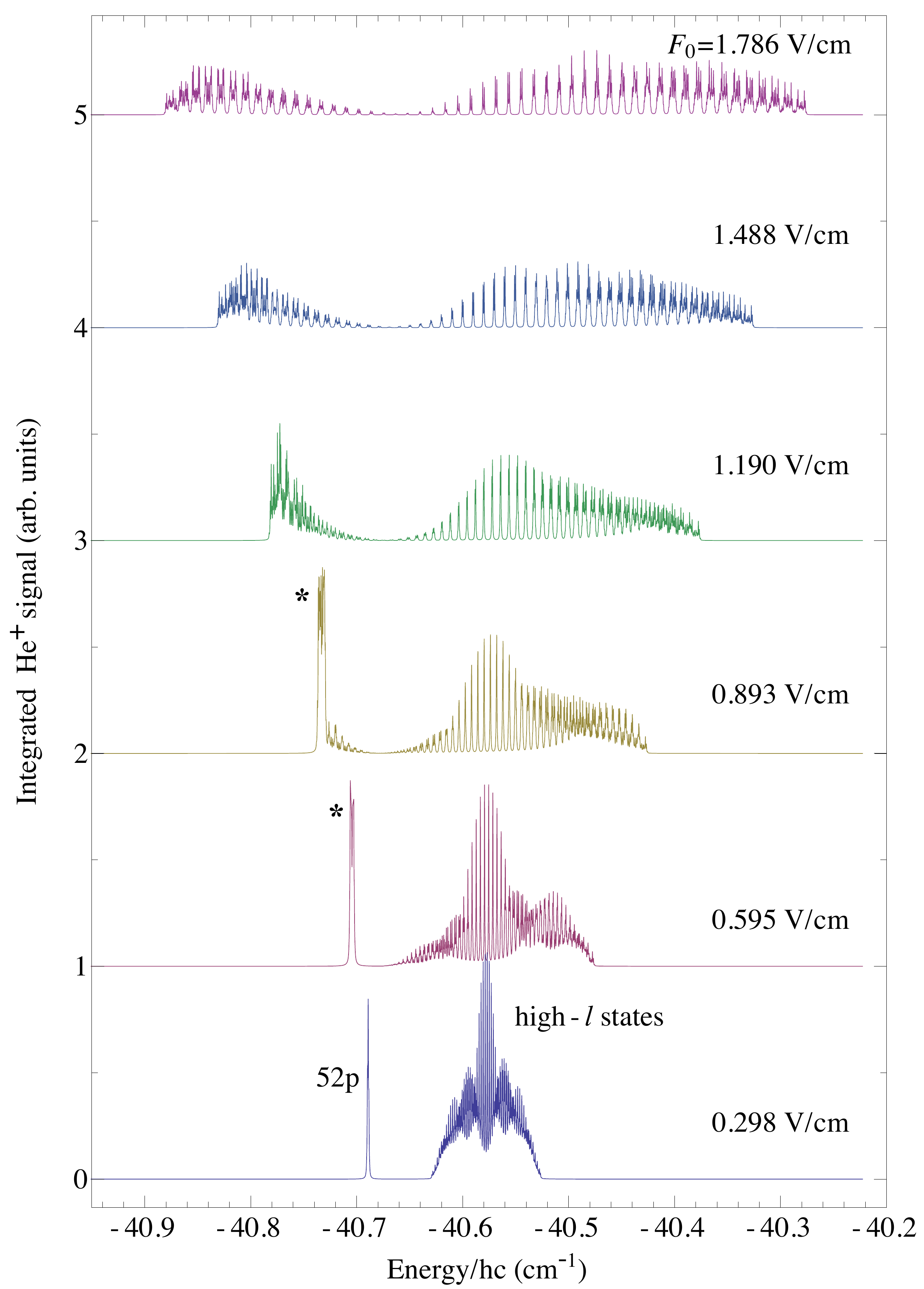}
\caption{Calculated spectra encompassing transitions to $n=52$ Rydberg states in electric field with dc offsets of $F_0=0.298$, 0.595, 0.893, 1.190 and 1.786~V/cm (bottom to top), and a modulation of amplitude $F_{\mathrm{osc}}=24.0$~mV/cm, and frequency $\omega_{\mathrm{rf}}=2\pi\times20$~MHz. The asterisks indicate the two calculated spectra for which the intensity of the transition to the field-free 52p state has been truncated to reflect the effects of saturation observed in the experiment (see text for details). For clarity of presentation the upper spectra are vertically offset.}
\label{fig6}
\end{center}
\end{figure}

An overview of the effects of amplitude modulation on the $n=52$ Rydberg-Stark states studied can be seen by comparing Figs.~\ref{fig3}--\ref{fig6}. These figures contain experimentally recorded and calculated spectra of 1s3p\,$^{3}$P$_2\rightarrow$ 1s52s/1s52d transitions in the presence of dc electric fields with magnitudes of $F_0=0.298, 0.595, 0.893, 1.190, 1.488$ and $1.786$~V/cm. Measurements and calculations without, and with, modulation are displayed in Fig.~\ref{fig3} and Fig.~\ref{fig4}, and Fig.~\ref{fig5} and Fig.~\ref{fig6}, respectively. The dc offset electric fields and spectral regions encompassed in each case are marked by the dashed vertical lines labeled i-vi in Fig.~\ref{fig2}. In the calculated spectra presented in Fig.~\ref{fig4} and Fig.~\ref{fig6}, the transition to each state in the Stark manifold is represented by a Lorentzian function with an amplitude corresponding to the calculated spectral intensity, and a full-width-at-half-maximum (FWHM) of 0.00043~cm$^{-1}$ ($\equiv13$~MHz) matching that of the experiment. 

In the bottom spectra in Fig.~\ref{fig3} and Fig.~\ref{fig4}, for which $F_0=0.298$~V/cm, the electric field introduces significant d-character to the isolated 52p state and mixes the higher-$\ell$ states giving rise to a manifold of Stark states with an approximately symmetric intensity distribution. This manifold of high-$\ell$ states is characterised by an energy splitting between neighbouring Stark states of $\sim0.00200$~cm$^{-1}$ ($\equiv$60~MHz). In this low field, the calculated spectrum is in good agreement with the experimental one, with the only noticeable difference relating to the spectral intensities in the centre of the manifold of high-$\ell$ states. We attribute the flattening of the centre states in the experimental spectrum to saturation of the excitation process. This observed saturation may contain many-body contributions as seen in other recent experiments~\cite{zhelyazkova15a}. As the value of $F_0$ is increased, the manifold of high-$\ell$ states becomes more spread out and in the highest field of $F_0=1.786$~V/cm the neighbouring states are separated by 0.01201~cm$^{-1}$ ($\equiv$360~MHz). Along with this increased Stark splitting, the 52p state gradually mixes completely with the higher-$\ell$ states. The intensity distribution across each spectrum reflects the s- and/or d- character of each Rydberg-Stark state. For the higher values of $F_0$, two interleaved sets of spectral lines can be distinguished. The set with the higher (lower) intensity corresponds to transitions to states for which $m_{\ell}=0$ ($|m_{\ell}|=1$).

All of the spectra in Fig.~\ref{fig3} and Fig.~\ref{fig4} are normalised such that the spectral intensities in the calculated spectrum at the highest field, $F_0=1.786$~V/cm, matches the experimentally measured intensities. We note that in the calculated spectra for $F_0=0.595$~V/cm and $F_0=0.893$~V/cm the intensity of the transition to the 52p state (indicated by the asterisks) has been truncated at $\sim$0.3 of its maximum value, reflecting the effects of saturation observed in the experiment. 

The spectra in Fig.~\ref{fig5} are analogous to those displayed in Fig.~\ref{fig3}, but in this case a modulation of amplitude $F_{\mathrm{osc}} = 24.0$~mV/cm and frequency $\omega_{\mathrm{rf}} = 2\pi\times20$~MHz is introduced to the applied electric field. The oscillatory component of the field most significantly affects states which possess non-zero electric dipole moments. The leftmost (lowest energy) state shown in each spectrum, which evolves adiabatically to the 52p state in zero electric field, and will be referred to as the field-free 52p state in the following, exhibits a quadratic Stark shift and for electric fields less than 0.3~V/cm has a small electric dipole moment, while for electric fields $F_0\gtrsim1$~V/cm it joins the manifold of the higher-$\ell$ states and exhibits a linear Stark shift and an electric dipole moment of $\sim10000$ D (see Fig.~\ref{fig2}). In the spectra recorded in the lowest offset field of $F_0=0.298$~V/cm, the field-free 52p state is unaffected by the modulation, as are the centre states in the Stark manifold which also possess zero electric dipole moments. Conversely, the outermost states of the Stark manifold have spectral intensity distributions which are strongly perturbed by the oscillating field as they possess electric dipole moments approaching $\sim10000$~D. The observed splitting of these outer Stark states occurs because the unperturbed spectral intensity of a single state becomes distributed over a number of Floquet sidebands. The number of Floquet sidebands populated is higher for states with larger electric dipole moments. For example, in a dc offset electric field of $F_0=0.595$~V/cm, the field-free 52p state possesses an electric dipole moment of approximately 2000~D and has its spectral intensity distributed over 9 Floquet sidebands ($-4\leq q\leq 4$). As the value of the dc offset field is increased, the field-free 52p state state acquires a larger electric dipole moment as it joins the manifold of higher-$\ell$ states, and is perturbed to a greater extent by the oscillatory component of the electric field. This is evident in the spectra recorded with $F_0=0.595$~V/cm and $F_0=0.893$~V/cm. For these values of the offset electric field, the field-free 52p state has dipole moments of approximately 4400~D and 7700~D, respectively, and the modulation causes the spectral intensity of this state to be distributed over several more sidebands.  A more detailed investigation reveals that for these two values of $F_0$, the state has its spectral intensity distributed over 11 and 13 Floquet sidebands, respectively. When the dc offset field is increased to $F_0=1.786$~V/cm, the field-free 52p state has joined the manifold of higher-$\ell$ states and now possesses an electric dipole moment which is increased to approximately 10000~D. As the energy spacing between the high-$\ell$ Stark states increases with increasing values of $F_0$, the effect of the modulation on the outermost states, which possess the largest electric dipole moments, becomes more easily observable. For a dc electric field of $F_0=1.786$~V/cm, the field-free 52p state, for example, has its spectral intensity distributed over 15 Floquet sidebands ($-7\leq q\leq 7$).

The effect of the oscillatory component of the electric field on the manifold of high-$\ell$ Stark states depends strongly on (i) the energy separation between neighbouring Stark states compared to the energy spacing between the Floquet sidebands, and (ii) the number of Floquet sidebands with $q=0$ character. The energy separation between neighbouring states of the high-$\ell$ manifold in the dc offset field of $F_0=0.298$~V/cm is approximately 0.00200~cm$^{-1}$ ($\equiv$60~MHz), bottom trace in Fig.~\ref{fig3} and Fig.~\ref{fig5}, while the separation between the Floquet sidebands is $2\pi\times20$~MHz. The electric dipole moment of the outermost Stark state from the high-$\ell$ manifold is approximately 10000~D which is sufficiently large to cause the spectral intensity of the state to be distributed over a number of Floquet sidebands. This means that in the presence of the oscillating field many of the Floquet sidebands of neighboring states overlap and are approximately degenerate. In addition, when $F_{\mathrm{osc}}=24.0$~mV/cm a significant number of sidebands (up to $-6\leq q\leq 6$ for the outermost states with the highest electric dipole moments) possess $q=0$ character. This is the reason why the spectrum recorded with amplitude modulation, and a dc offset of $F_0=0.298$~V/cm is most significantly modified when compared to the corresponding spectrum in Fig.~\ref{fig3}. Note that in the calculated spectra for which $F_0=0.298$~V/cm and $F_0=0.893$~V/cm the states in the middle of the Stark manifold have higher amplitudes than observed experimentally. As in the spectra in Fig.~\ref{fig3}, this is attributed to effects of saturation.

\subsection{An isolated Rydberg-Stark state in amplitude-modulated electric fields}
\label{subsec:isolated}
\begin{figure}[!h]
\begin{center}
\includegraphics[width=\textwidth]{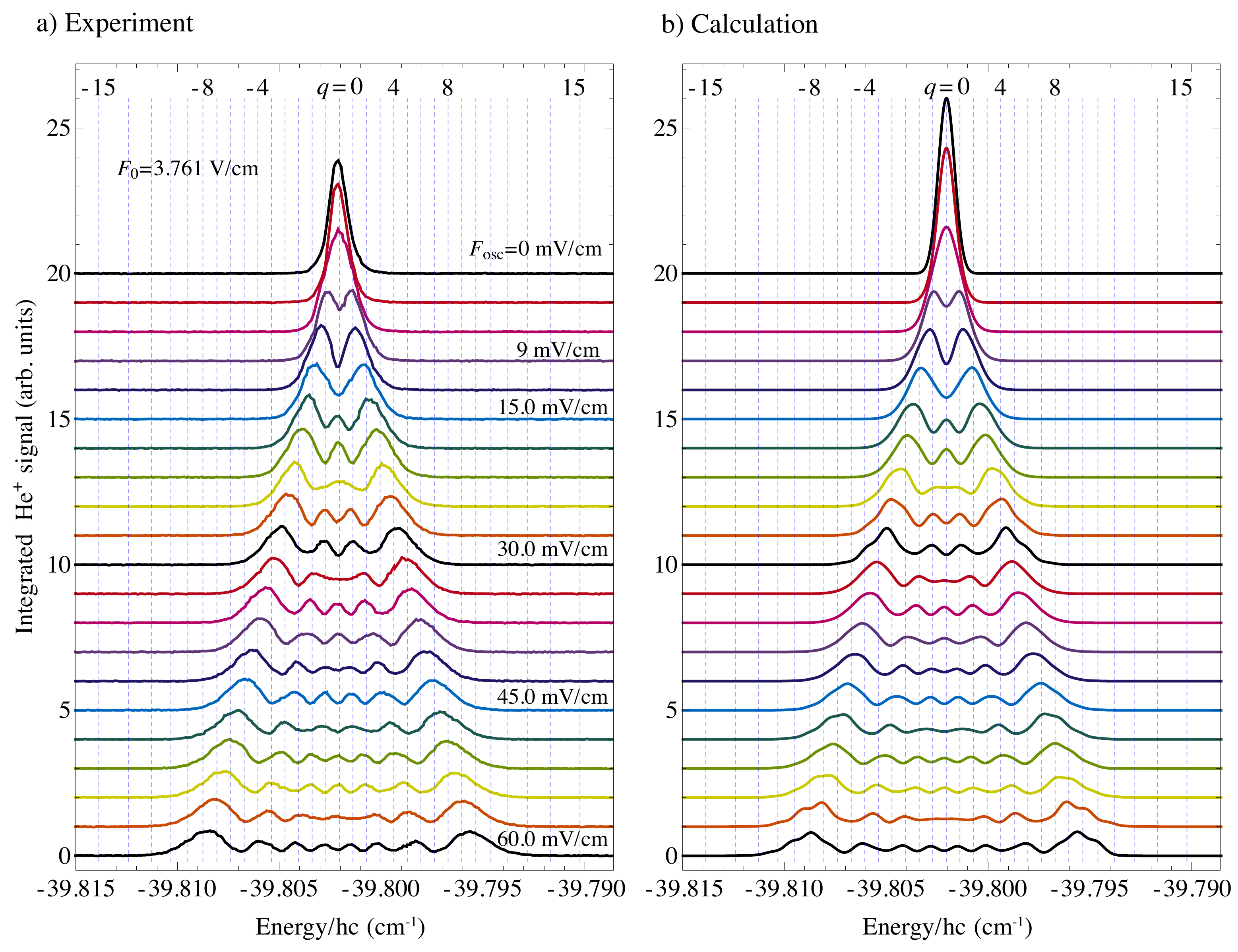}
\caption{(a) Experimentally recorded, and (b) calculated  spectra encompassing the spectral region indicated by the dot labelled $A$ in Fig.\ref{fig1}. The spectra were recorded in an electric field with a dc offset of $F_0=3.761$~V/cm, and modulation at a frequency of $\omega_{\mathrm{rf}}=2\pi\times20$~MHz. The modulation amplitude was adjusted from $F_{\mathrm{osc}}=0$~mV/cm (top trace) to $F_{\mathrm{osc}}=60.0$~mV/cm (bottom trace) in steps of 3~mV/cm. The Floquet sidebands are shown as dashed vertical lines.}
\label{fig7}
\end{center}
\end{figure}

To investigate in more detail the effect of the oscillating electric field on an isolated Rydberg-Stark state which has a large electric dipole moment, and to provide a more quantitative comparison of the experimental data with the Floquet calculations, we have studied the outermost, lowest  energy Stark state of the $n=53$ Stark manifold which evolves adiabatically to the 53s state in zero electric field. In an electric field of $F_0=3.760$~V/cm (red dot labelled $A$ in Fig.~\ref{fig2}) there are no other Stark states in the immediate vicinity of this state and it has an electric dipole moment of approximately 7900 D. The spectra shown in Fig.~\ref{fig7}(a) were recorded as the wavenumber of the ir laser was scanned over 0.020~cm$^{-1}$ around the 1s3p $^3$P$_2\rightarrow1$s53s transition in this field with modulation at a frequency $\omega_{\mathrm{rf}}=2\pi\times20$~MHz. The modulation amplitude was adjusted from $F_{\mathrm{osc}}=0$~mV/cm (top curve) to $F_{\mathrm{osc}}=60.0$~mV/cm (bottom curve) in steps of 3~mV/cm. Without modulation, the spectral feature has a Gaussian shape (top traces in Fig.~\ref{fig7}). For small amplitude modulation (e.g., $F_{\mathrm{osc}}=3$~mV/cm and 6~mV/cm), the height of the Gaussian reduces slightly, while its width increases. As the amplitude of the modulation is increased further, the spectral intensity distribution becomes more significantly modified. This modification can be understood with the aid of the Floquet calculations. Displayed in Fig.~\ref{fig7}(b) are the calculated spectral intensity distributions corresponding to the experimental data shown in Fig.~\ref{fig7}(a). Each spectrum in Fig.~\ref{fig7}(b) was obtained by convoluting the calculated spectral intensities with Gaussian functions with FWHM of 0.00093~cm$^{-1}$ ($\equiv$28~MHz). The slightly increased spectral width of this transition compared to those in Fig.~\ref{fig4} and Fig.~\ref{fig6} arises from a combination of its extreme sensitivity to rf electrical laboratory noise, its intensity, and dipolar interactions within the atomic beam~\cite{zhelyazkova15a}. Also shown in Fig.~\ref{fig7} as dashed vertical lines are 31 Floquet sidebands, $-15\leq q\leq15$, more of which acquire $q=0$ character as the amplitude of the modulation is increased. For a modulation amplitude of $F_{\mathrm{osc}}=9$~mV/cm [fourth trace from the top in Fig.~\ref{fig7}(a) and (b)], the unperturbed spectral intensity is distributed between the $q=\pm1$ sidebands, with a reduction of the intensity of the $q=0$ sideband. The number of sidebands observed in the spectra is a consequence of the amplitude of the Stark energy shift of the state in the oscillating field. As $F_{\mathrm{osc}}$ is increased, the corresponding Stark energy shifts of the Rydberg state increase, and Floquet sidebands which are further displaced from the unperturbed energy level attain $q=0$ character and hence spectral intensity.

The complete spectral intensity distribution depends on the $q=0$ character of each Floquet sideband. It can be clearly seen that the ratio of the intensity of the $q=0$ sideband to the intensities of the $q=\pm1$ sidebands oscillates as $F_{\mathrm{osc}}$ is increased, and is particularly sensitive to the value of $F_{\mathrm{osc}}$. In all of the data presented in Fig.~\ref{fig7}, there is excellent quantitative agreement between the experimental spectra and the results of the calculations. This was achieved upon including sidebands up to $q_{\mathrm{max}}=17$ in the calculation. Comparison of the results of the calculations with the experimental data indicates a sensitivity in the measurements to changes in $F_{\mathrm{osc}}$ of $\pm500$~$\mu$V/cm.

\subsection{Effects of amplitude modulation at avoided crossings}
\begin{figure}[!h]
\begin{center}
\includegraphics[width=\textwidth]{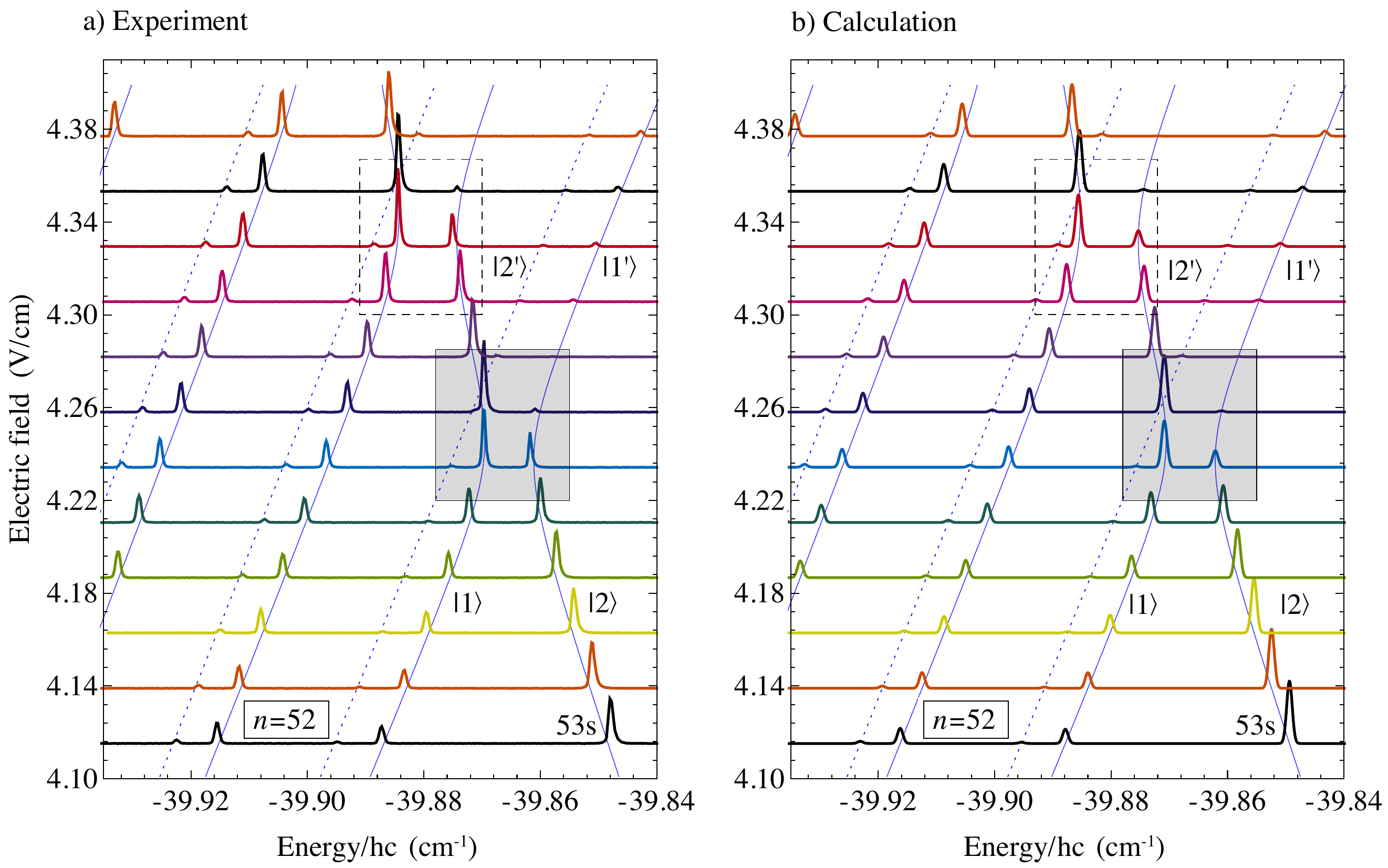}
\caption{(a) Experimentally recorded, and (b) calculated spectra encompassing transitions to Stark states near the Inglis-Teller limit between the $n=52$ and $n=53$ Stark manifolds. These spectra were recorded for dc offset electric fields ranging from $F_0=4.115$~V/cm to 4.377~V/cm in steps of 0.022~V/cm (as indicated on the vertical axes), without modulation. The solid (dashed) curves show the calculated energies of the Stark states with $m_{\ell}=0$ ($|m_{\ell}|$=1). The shaded gray rectangles indicate the region at the avoided crossing involving states $|1\rangle$ and $|2\rangle$ discussed in the text, while the dashed rectangles indicate the avoided crossing between state $|2'\rangle$ and the second outermost $m_{\ell}=0$ Stark state from the $n=52$ manifold.}
\label{fig8}
\end{center}
\end{figure}

To test the limits of the Floquet calculations, we investigate the changes in spectral intensity distributions of two states which interact at an avoided crossing at the Inglis-Teller limit between the $n=52$ and $n=53$ Stark manifolds, in the presence of a time-dependent electric field. The two states of interest are: state $|1\rangle$~- the outermost, highest-energy (low-field-seeking) Stark state of the $n=52$ Stark manifold, and state $|2\rangle$~- the outermost, lowest-energy (high-field-seeking) Stark state of the $n=53$ Stark manifold, which evolves adiabatically to the 53s state in zero electric field. In the case of an adiabatic (diabatic) traversal of the crossing state $|1\rangle$ evolves into state $|2'\rangle$ ($|1'\rangle$) and state $|2\rangle$ evolves into state $|1'\rangle$ ($|2'\rangle$) after the crossing as indicated in Fig.~\ref{fig8}.

Experimentally recorded and calculated spectra encompassing the transitions to several of the outermost Stark states in the spectral region surrounding this avoided crossing in dc offset electric fields ranging from $F_{0}=4.115$~V/cm to $F_{0}=4.377$~V/cm in steps of 0.022~V/cm are displayed in Fig.~\ref{fig8}. This spectral region is indicated by the box labelled $B$ in the Stark map in Fig.~\ref{fig2}. As the value of the dc offset field is increased from $F_0=4.115$~V/cm to $F_0=4.234$~V/cm, the transition to the state $|2\rangle$ gradually reduces in intensity, while the transition to state $|1\rangle$ increases in intensity. Beyond the avoided crossing, for dc offset fields from $F_0=4.234$~V/cm to $F_0=4.380$~V/cm, state $|1'\rangle$ loses most of its intensity with the transition only just observable, while state $|2'\rangle$ retains its intensity. This behaviour characterizes the adiabatic evolution to states $|2'\rangle$ and $|1'\rangle$. Two sets of transitions are visible in the spectra displayed in Fig.~\ref{fig8}. The set with the higher (lower) intensity corresponds to transitions to states for which $m_{\ell}=0$ ($|m_{\ell}|=1$). These relative intensities are similar to those in the spectra presented in Fig.~\ref{fig3}. Also shown in Fig.~\ref{fig8} as solid (dashed) curves are the calculated Stark energies of the $m_{\ell}=0$ ($|m_{\ell}|=1$) states, from which it can be seen how the states with $m_{\ell}=0$ ($|m_{\ell}|=1$) shift at the crossing. The differences exhibited by these two sets of transitions at the crossing are the result  of the differences in magnitude of the quantum defects for the s- and d-states. The $m_{\ell}=0$ component of state $|1\rangle$ interacts strongly with the $m_{\ell}=0$ component of state $|2\rangle$ because the quantum defect for s-states is large ($\delta_{s}=0.296671$ for $n=52$). The Stark states from the $n=52$ and $n=53$ manifolds for which $|m_{\ell}|=1$, on the other hand, interact with each other to a lesser degree, and thus almost cross exactly, because they are only composed of states with $\ell\geq1$, for which $\delta_{\ell}\leq 0.068353$. Also visible in Fig.~\ref{fig8} are the avoided crossings between the $m_{\ell}=0$ component of the second outermost Stark state of the $n=52$ Stark manifold and state $|2'\rangle$ (dashed rectangles in Fig.~\ref{fig8}).      
   
\begin{figure}[!h]
\includegraphics[width=\textwidth]{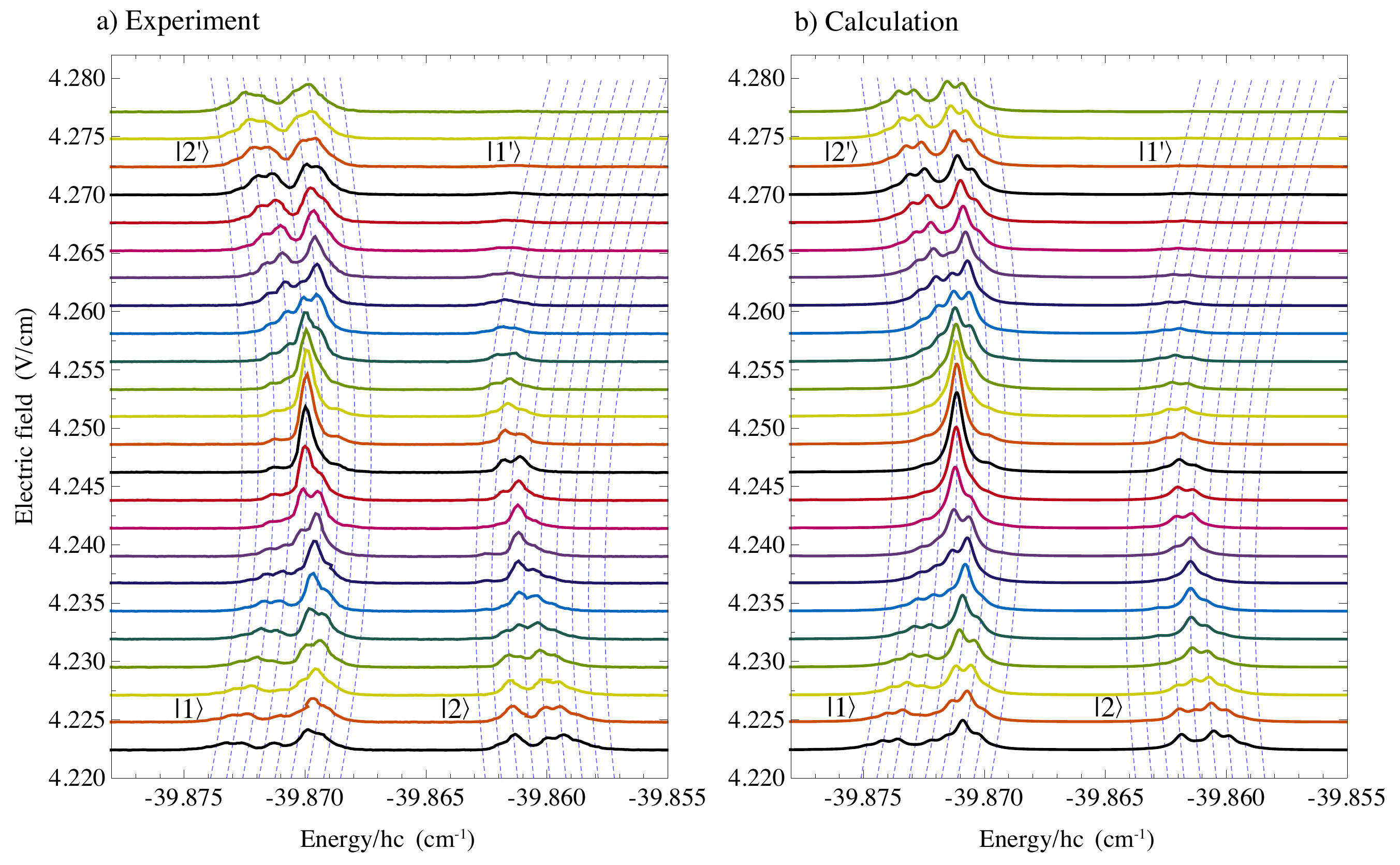}
\caption{(a) Experimentally recorded, and (b) calculated spectra encompassing transitions to states $|1\rangle$ and $|2\rangle$ (see text and Fig.~\ref{fig8}) in the presence of time-varying electric fields with dc offsets ranging from $F_0=4.222$~V/cm to $F_0=4.277$~V/cm in steps of 2~mV/cm. For all spectra modulation was applied with an amplitude $F_{\mathrm{osc}}=24.0$~mV/cm, and frequency $\omega_{\mathrm{rf}}=2\pi\times20$~MHz. Also shown as dashed curves are 9 Floquet sidebands associated with each state.}
\label{fig9}
\end{figure}

To demonstrate the effects of amplitude modulation at an avoided crossing, we show in Fig.~\ref{fig9} experimentally recorded and calculated spectra encompassing transitions to the states $|1\rangle$, $|2\rangle$, $|1'\rangle$ and $|2'\rangle$ in electric fields with dc offsets ranging from $F_0=4.222$~V/cm to $F_0=4.278$~V/cm in steps of 2~mV/cm, and modulation with an amplitude $F_{\mathrm{osc}}=24.0$~mV/cm, and a frequency $\omega_{\mathrm{rf}}=2\pi\times20$~MHz. The dc offset fields and wavenumber ranges corresponding to the region of interest are indicated by the shaded rectangles in Fig.~\ref{fig8}. For the smallest dc electric field of $F_0=4.222$~V/cm (bottom traces in Fig.~\ref{fig9}) states $|1\rangle$ and $|2\rangle$ have electric dipole moments of 6000~D and 3600~D, respectively, and their spectral intensity is therefore distributed among several Floquet sidebands as a result of the modulation. In these cases $\sim9$ Floquet sidebands (dashed curves in Fig.~\ref{fig9}) are populated. As the value of the dc offset electric field is increased from $F_0=4.222$~V/cm to 4.244~V/cm, both states lose their electric dipole moments as they begin to approach the avoided crossing, and their spectral intensity distributions are less affected by the modulation. For dc fields from $4.244$~V/cm to $4.254$~V/cm, between which lies the avoided crossing, state $|1\rangle$ has a spectral distribution of maximum intensity and is characterized by a single Lorentzian function, while the 53s state has almost completely lost its intensity. The single intensity maximum is characteristic of a state with zero electric dipole moment as occurs exactly at a crossing point. As the dc electric field is increased from $F_0=4.254$~V/cm to 4.278~V/cm, states $|1'\rangle$ and $|2'\rangle$ start to acquire non-zero electric dipole moments again, reaching values of 8100~D and 4700~D, respectively, for $F_0=4.278$~V/cm. Correspondingly, their spectral intensity distributions are split into an increasing number of components.   

\begin{figure}[!h]
\includegraphics[width=\textwidth]{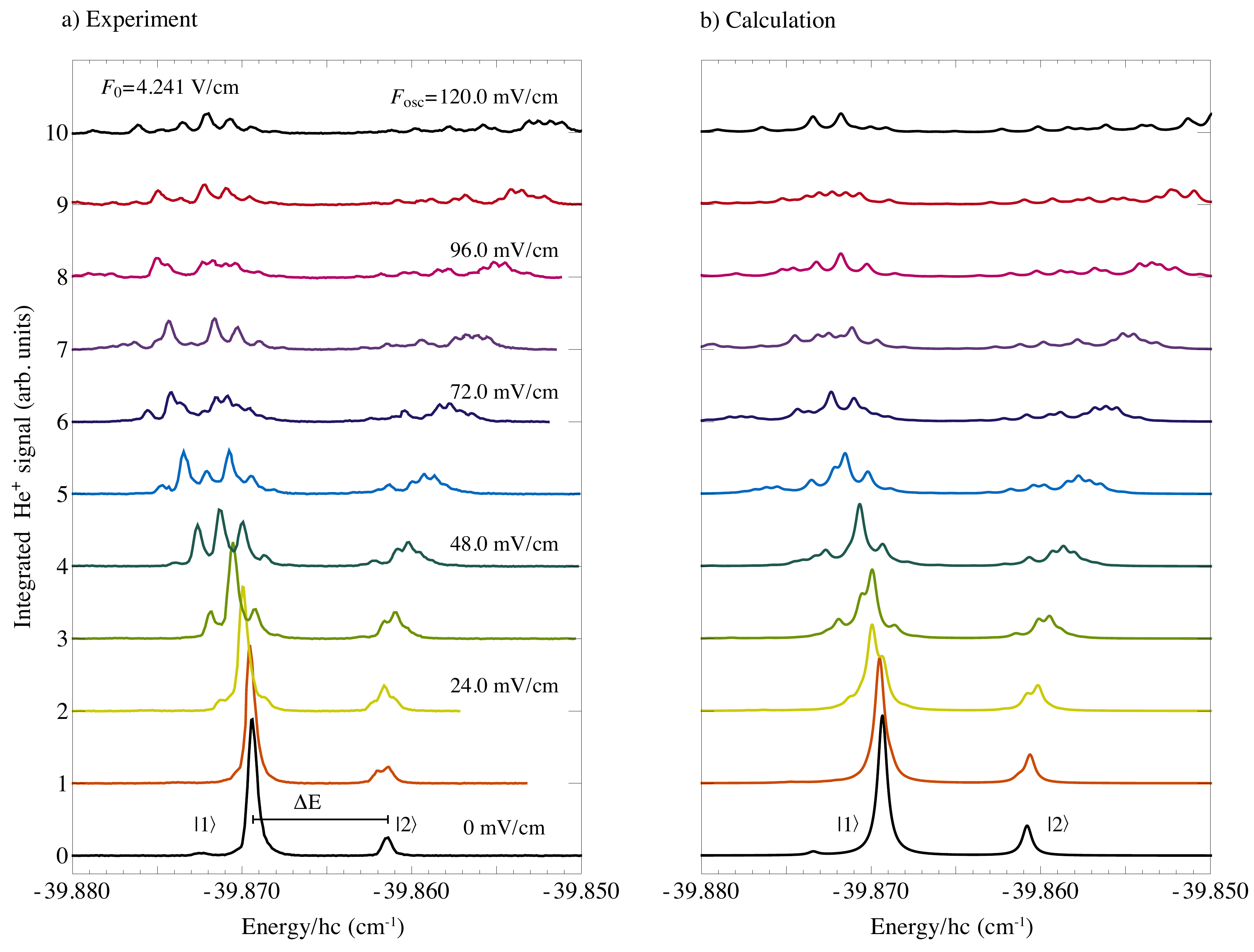}
\caption{(a) Experimentally recorded, and (b) calculated spectra encompassing the transitions to the outermost, highest-energy Stark state of the $n=52$ Stark manifold (state $|1\rangle$) and the outermost, lowest-energy Stark state of the $n=53$ Stark manifold (state $|2\rangle$). In all spectra a dc offset electric field of $F_0=4.241$~V/cm was applied, with modulation at a frequency $\omega_{\mathrm{rf}}=2\pi\times20$~MHz. The modulation amplitude, $F_{\mathrm{osc}}$, was adjusted from 0~V/cm to 120~mV/cm in steps of 12.0~mV/cm as indicated.}
\label{fig10}
\end{figure}

The agreement between the experimentally recorded and calculated spectra in Fig.~\ref{fig9} suggests that the weak modulation of amplitude $F_{\mathrm{osc}}=24.0$~mV/cm, and frequency $\omega_{\mathrm{rf}}=2\pi\times20$~MHz does not give rise to significant effects not accounted for in the Floquet calculations. However, to test the validity of the Floquet calculations in fields with stronger modulation, we investigated the effects of a modulation at a fixed frequency, but increasing amplitudes, $F_{\mathrm{osc}}$, exactly at the avoided crossing. The results of the experiments and calculations at the crossing point between states $|1\rangle$ and $|2\rangle$, for an electric field with a fixed dc offset of $F_0=4.241$~V/cm, and modulation at a frequency $\omega_{\mathrm{rf}}=2\pi\times20$~MHz and amplitude varied from $F_{\mathrm{osc}}=0$~V/cm to 120~mV/cm in steps of $\sim12$~mV/cm, are displayed in Fig.~\ref{fig10}. 

\begin{figure}[!h]
\begin{center}
\includegraphics[width=0.6\textwidth]{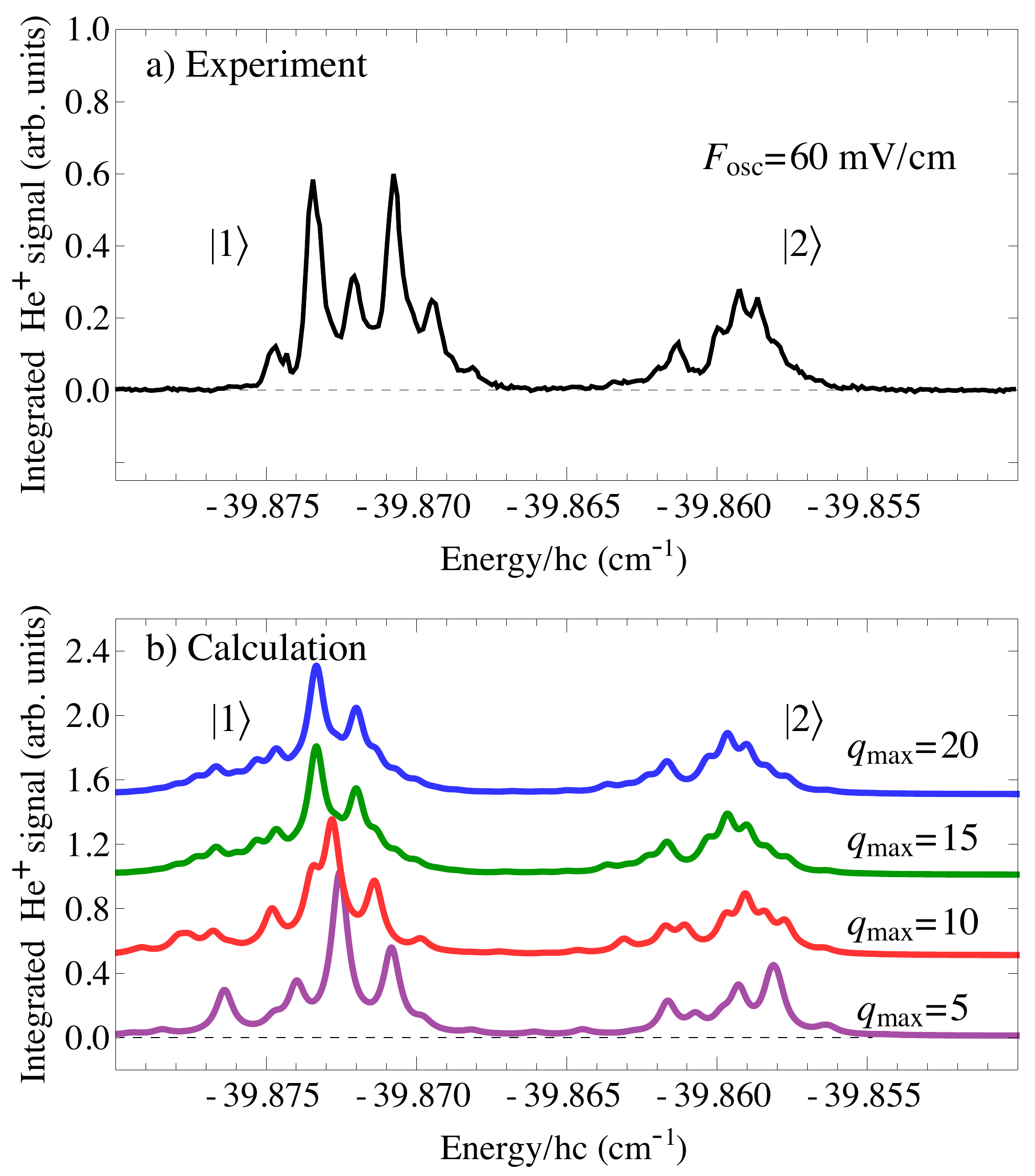}
\caption{(a) Experimentally recorded spectrum in a dc offset electric field of $F_0=4.241$~V/cm, with modulation at a frequency $\omega_{\mathrm{rf}}=2\pi\times20$~MHz and amplitude $F_{\mathrm{osc}}=60$~mV/cm [as in the sixth spectrum from the bottom in Fig.~\ref{fig10}(a)]. (b) Calculated spectra for the same spectral region and values of $F_0$, $F_{\mathrm{osc}}$ and $\omega_{\mathrm{rf}}$ as in (a) but with basis sets for which $q_{\mathrm{max}}=5$--$20$ as indicated.}
\label{fig11}
\end{center}
\end{figure}

The data presented in Fig.~\ref{fig10} shows that there is good agreement between the experimentally recorded and calculated spectra for values of $F_{\mathrm{osc}}$ below 36~mV/cm. For modulation amplitudes larger than this increasing deviations of the calculated spectra from the experimental data are seen. To confirm that the calculations presented in Fig.~\ref{fig10}(b) have converged and a sufficiently large number of sidebands have been included, the spectrum for which $F_{\mathrm{osc}}=60$~mV/cm was studied in more detail (see Fig.~\ref{fig11}). Comparison of calculations for which $q_{\mathrm{max}}=5, 10, 15$ and 20 [Fig.~\ref{fig11}(b)] with the experimental data [Fig.~\ref{fig11}(a)] reveals that for values of $q_{\mathrm{max}}>15$ identical spectra are obtained, confirming that convergence has been achieved. However, there remains a considerable discrepancy between the calculation and the experimental spectrum [Fig.~\ref{fig11}(a)]. This disparity is most notable on the low-energy side of state $|1\rangle$. 

The link between the classical electron dynamics in the amplitude-modulated fields at the avoided crossing, with those of an oscillating linear quadrupole suggests that the deviations between the experiments and calculated spectra may result from higher-order interactions, beyond the electric dipole couplings considered in the Floquet calculations. This, together with contributions from the high local density of excited states on the photoexcitation dynamics are not accounted for in the calculation presented here. Studies of these effects for modulation at a larger range of frequencies and amplitudes will be the subject of future experiments and theoretical work. 

\section{Conclusion}

The experiments and calculations reported here demonstrate the sensitivity of high Rydberg states to low-frequency oscillating electric fields. For states with electric dipole moments of $\sim10000$~D, comparison of the experimental data with the results of the Floquet calculations indicates that the measurements are sensitive to changes of $\pm500~\mu$V/cm in the amplitude of the electric field modulation, for a frequency of 20~MHz. Rydberg-Stark states therefore represent excellent microscopic probes of low-frequency electrical noise, causing minimal perturbations to the local electrical environment. This sensitivity may be of particular value in the identification of electrical noise close to surfaces which can lead to decoherence of quantum systems, and is also of importance in Rydberg atom/molecule scattering experiments preformed at vacuum--solid-state interfaces. The work described here also highlights effects that may arise as a result of the effective oscillating electric fields experienced by Rydberg atoms and molecules confined in electrostatic traps, and the oscillating electric fields experienced by ions in high Rydberg states trapped in Paul traps~\cite{feldker15}. It is therefore expected to be of importance in the interpretation and refinement of experiments in which atoms and molecules in Rydberg-Stark states with large electric dipole moments are decelerated, transported and trapped using inhomogeneous electric fields.

\section{Acknowledgements}
This work was supported financially by the Department of Physics and Astronomy and the Faculty of Mathematical and Physical Sciences at University College London, and the Engineering and Physical Sciences Research Council under Grant No. EP/L019620/1.

\label{lastpage}


\begin{thebibliography}{9}

\bibitem{dethlefs84a} K. M\"uller-Dethlefs, M. Sander, E. W. Schlag, Chem. Phys. Lett. {\bf{112}}, 291 (1984).

\bibitem{reiser88a} G. Reiser, W. Habenicht, K. M\"uller-Dethlefs, E. W. Schlag, Chem. Phys. Lett. {\bf{152}}, 119 (1988).

\bibitem{hollenstein01a}  U. Hollenstein, R. Seiler, H. Schmutz, M. Andrist and F. Merkt, J. Chem. Phys. {\bf{115}}, 5461 (2001).

\bibitem{liu09a} J. Liu, E. J. Salumbides, U. Hollenstein, J. C. J. Koelemeij, K. S. E. Eikema, W. Ubachs, and F. Merkt, J. Chem. Phys. {\bf{130}}, 174306 (2009).

\bibitem{sprecher11a} D. Sprecher, Ch. Jungen, W. Ubachs and F. Merkt, Faraday Discuss. {\bf{150}}, 51 (2011).

\bibitem{dunning84a} F. B. Dunning and R. F. Stebbings, Ann. Rev. Phys. Chem. {\bf{33}}, 173 (1982).

\bibitem{hill00a} S. B. Hill, C. B. Haich, Z. Zhou, P. Nordlander, and F. B. Dunning, Phys. Rev. Lett. {\bf{85}}, 5444 (2000).

\bibitem{lloyd05a} G. R. Lloyd, S. R. Procter, and T. P. Softley, Phys. Rev. Lett. {\bf{95}}, 133202 (2005).

\bibitem{sashikesh13a} G. Sashikesh, M. S. Ford, and T. P. Softley, J. Chem. Phys. {\bf{138}}, 114308 (2013).

\bibitem{yamakita04a}  Y. Yamakita, S. R. Procter, A. L. Goodgame, T. P. Softley and F. Merkt, J. Chem. Phys. {\bf{121}}, 1419 (2004).

\bibitem{vliegen04a} E. Vliegen, H. J. W\"orner, T. P. Softley and F. Merkt, Phys. Rev. Lett. {\bf{92}}, 033005 (2004).

\bibitem{hogan12a} S. D. Hogan, P. Allmendinger, H. Sa\ss mannshausen, H. Schmutz and F. Merkt, Phys. Rev. Lett.  {\bf{108}}, 063004 (2012).

\bibitem{lancuba14a} P. Lancuba and S. D. Hogan, Phys. Rev. A {\bf{90}}, 053420 (2014).

\bibitem{lancuba13a} P. Lancuba and S. D. Hogan, Phys. Rev. A {\bf{88}}, 043427 (2013).

\bibitem{allmendinger14a} P. Allmendinger, J. Deiglmayr, J. A. Agner, H. Schmutz, and F. Merkt, Phys. Rev. A {\bf{90}}, 043403 (2014).

\bibitem{ko14a} H. Ko and S. D. Hogan, Phys. Rev. A {\bf{89}}, 053410 (2014).

\bibitem{hogan08a} S. D. Hogan and F. Merkt, Phys. Rev. Lett. {\bf{100}}, 043001 (2008).

\bibitem{hogan09a} S. D. Hogan, Ch. Seiler, and F. Merkt, Phys. Rev. Lett. {\bf{103}}, 123001(2009).

\bibitem{seiler11a} Ch. Seiler, S. D. Hogan, and F. Merkt, Phys. Chem. Chem. Phys. {\bf{13}}, 19000 (2011).

\bibitem{seiler11b} Ch. Seiler, S. D. Hogan, H. Schmutz, J. A. Agner and F. Merkt, Phys. Rev. Lett. {\bf{106}}, 073003 (2011).

\bibitem{osterwalder99a} A. Osterwalder and F. Merkt, Phys. Rev. Lett. {\bf{82}}, 1831 (1999).

\bibitem{autler56a} S. H. Autler and C. H. Townes, Phys. Rev. {\bf{100}}, 703 (1956).

\bibitem{gallagher08a} T. F. Gallagher and P. Pillet, Adv. At. Mol. Opt. Phys. {\bf{56}}, 161 (2008).

\bibitem{pillet83} P. Pillet, R. Kachru, N. H. Tran, W. W. Smith, and T. F. Gallagher, Phys. Rev. Lett. {\bf{50}}, 1763 (1983).

\bibitem{vandenHeuvell84} H. B. van Linden van den Heuvell, R. Kachru, N. H. Tran, and T. F. Gallagher, Phys. Rev. Lett. {\bf 53}, 1901 (1984).  

\bibitem{pillet87} P. Pillet, R. Kachru, N. H. Tran, W. W. Smith, and T. F. Gallagher, Phys. Rev. A {\bf 36} 1132 (1987).

\bibitem{zhang94a} Y. Zhang, M. Ciocca, L.-W. He, C. E. Burkhardt, and J. J. Leventhal, Phys. Rev. A {\bf{50}}, 1101 (1994).

\bibitem{spellmeyer97a} N. W. Spellmeyer, D. Kleppner, M. R. Haggerty, V. Kondratovich,
J. B. Delos, and J. Gao, Phys. Rev. Lett. {\bf{79}}, 1650 (1997).

\bibitem{yoshida12a} S. Yoshida, C. O. Reinhold, J. Burgd\"orfer, S. Ye, and F. B. Dunning, Phys. Rev. A {\bf{86}}, 043415 (2012).

\bibitem{vanditzhuijzen09a} C. S. E. van Ditzhuijzen, A. Tauschinsky, and H. B. van Linden van den Heuvell, Phys. Rev. A {\bf{80}}, 063407 (2009).

\bibitem{tretyakov14a} D. B. Tretyakov, V. M. Entin, E. A. Yakshina, I. I. Beterov, C. Andreeva, and I. I. Ryabtsev, Phys. Rev. A {\bf{90}}, 041403(R) (2014).

\bibitem{zhelyazkova15a} V. Zhelyazkova and S. D. Hogan, Phys. Rev. A {\bf{92}}, 011402(R) (2015).

\bibitem{gallagher94a} T. F. Gallagher, \emph{Rydberg Atoms} (Cambridge University
Press, Cambridge, 1994).

\bibitem{shirley65a} J. H. Shirley, Phys. Rev. {\bf{138}}, 979 (1965).

\bibitem{halfmann00a} T. Halfmann, J. Koensgen, and K. Bergmann, Meas. Sci. Technol. {\bf{11}}, 1510 (2000).

\bibitem{zimmerman79a} M. L. Zimmerman, M. G. Littman, M. M. Kash, and D. Kleppner, Phys. Rev. A {\bf{20}}, 2251 (1979).

\bibitem{grimmel15} J. Grimmel, M. Mack, F. Karlewski, F. Jessen, M. Reinschmidt, N. S\'{a}ndor and J. Fort\'{a}gh, N. J. Phys. {\bf{17}}, 053005 (2015). 

\bibitem{drake91} G. W. F. Drake and R. A. Swainson, Phys. Rev. A {\bf{44}}, 5448 (1991).

\bibitem{spellmeyer98a} N. W. Spellmeyer, Ph. D. thesis, Massachusetts Institute of Technology (1998).

\bibitem{zare88a} R. N. Zare, \emph{Angular Momentum}, (John Wiley \& Sons, New York, 1988).

\bibitem{feldker15} T. Feldker, P. Bachor, M. Stappel, D. Kolbe, R. Gerritsma, J. Walz, and F. Schmidt-Kaler, Phys. Rev. Lett. {\bf{115}}, 173001 (2015).

\end{thebibliography}
\end{document}